\documentclass[12pt]{article}

\usepackage{geometry}
\usepackage{relsize}
\usepackage[onehalfspacing]{setspace}
\usepackage{parskip}

\usepackage{cite}

\usepackage[english]{babel}
\usepackage{amsmath}
\usepackage{amssymb}

\newcommand{\dd}{\mbox{d}}

\usepackage[colorlinks=true,citecolor=green,linkcolor=red,filecolor=cyan,urlcolor=magenta]{hyperref}

\begin{document}

\title{Black hole entropy contributions from Euclidean cores}

\author{Jens Boos\,\footnote{~E-mail: \href{mailto:jboos@wm.edu}{jboos@wm.edu}} \\
{\small High Energy Theory Group, Department of Physics, William \& Mary}\\[-8pt]
{\small Williamsburg, VA 23187-8795, United States} }

\date{May 11, 2023}

\maketitle

\begin{abstract}

The entropy of a Schwarzschild black hole, as computed via the semiclassical Euclidean path integral in a stationary phase approximation, is determined not by the on-shell value of the action (which vanishes), but by the Gibbons--Hawking--York boundary term evaluated on a suitable hypersurface, which can be chosen arbitrarily far away from the horizon. For this reason, the black  hole singularity seemingly has no influence on the Bekenstein--Hawking area law. In this Essay we estimate how a regular black hole core, deep inside a Euclidean black hole of mass $M$ and generated via a UV regulator length scale $\ell > 0$, affects the black hole entropy. The contributions are suppressed by factors of $\ell/(2GM)$; demanding exact agreement with the area law as well as a self-consistent first law of black hole thermodynamics at all orders, however, demands that these contributions vanish identically via uniformly bounded curvature. This links the limiting curvature hypothesis to black hole thermodynamics.


{\hfill \smaller \textit{file: euclidean-cores-v5.tex, May 11, 2023, jb} }

\vspace{2cm}
\begin{center}
{\small Essay written for the Gravity Research Foundation 2023 Awards for Essays on Gravitation; awarded an Honorable Mention.}
\end{center}

\end{abstract}

\vfill

\pagebreak

\section{Introduction}

In their seminal 1976 paper, Gibbons and Hawking \cite{Gibbons:1976ue} demonstrated how to link the semiclassical thermodynamics of vacuum black holes to the on-shell value of the gravitational action. Nowadays, more than 300 papers per year are devoted to the subject of black hole thermodynamics. The thermodynamic behavior of black holes is rich, and often surprising, and has given rise to entirely new subdisciplines, such as ``black hole chemistry'' \cite{Kubiznak:2014zwa}. The fact that black holes radiate, and eventually evaporate completely, is also important for the information loss problem of black holes. What is the final stage of black hole evaporation? And can it help us to address fundamental questions in quantum gravity?

In this Essay, we would like to ask (and tentatively answer) a simple question: how does the black hole interior contribute to black hole thermodynamics, and, in particular, to the black hole entropy? After all, black holes in Einstein gravity contain essential singularities of unbounded gravitational fields, but thermodynamics is seemingly unfazed by those strong fields. Conversely, considerable effort has been devoted to the resolution of classical black hole singularities both from basic principles as well as in proposed non-singular black hole geometries. The study of a latter has become an active field of research \cite{Frolov:2016pav}, but the inherent ambiguity of regularizing procedures has made it difficult to favor one approach over the other.

As we will show, a handful of reasonably simple assumptions about the black hole interior is enough to estimate the general form of black hole entropy corrections. Moreover, as we will see, at leading order consistent thermodynamics is compatible with a mass-dependent, ``dressed'' regulator
\begin{align}
\ell \sim \frac{\ell_\text{Pl}^2}{2 G M} \, ,
\end{align}
where $M$ is the mass of the black hole under consideration. These are our assumptions:
\begin{enumerate}
\item[A.] In a suitable extension of General Relativity, there exists a vacuum black hole solution that features a regular, horizonless Euclidean core.\\[-1.4\baselineskip]
\item[B.] For astrophysical black holes described by such metrics, their Hawking temperature and near-horizon geometry are essentially identical to those of General Relativity.\\[-1.4\baselineskip]
\item[C.] In these solutions, the new physics are parametrized via a length scale $\ell > 0$, which may or may not coincide with the Planck scale $\ell_\text{Pl}$. The presence of this length scale regulates divergences encountered in General Relativity, and for $\ell \rightarrow 0$ we recover General Relativity.
\end{enumerate}
With the assumptions stated, let us now explore what we can estimate based on them. We will work in units where $\hbar = c = k_\text{B} = 1$, make use of the convention $G = \ell_\text{Pl}^2$, and, to avoid confusion, we will denote the entropy as ``$S$'' and the action as ``$I$.''

\section{Estimating black hole entropy corrections}
The total gravitational action $I_\text{tot}$, as a sum of the Einstein--Hilbert action as well as the  Gibbons--Hawking--York boundary term, is given by
\begin{align}
I_\text{tot} = I_\text{EH} + I_\text{GHY} = -\frac{1}{16\pi G} \int\limits_{\mathcal{M}} \dd^4 X \sqrt{-g} R - \frac{1}{8\pi G} \int\limits_{\partial\mathcal{M}} \dd^3 x \sqrt{-g} K \, ,
\end{align}
where $\mathcal{M}$ is the domain of integration (with coordinates $X{}^\mu$), $\partial\mathcal{M}$ is its boundary (with coordinates $x{}^{\mu}$), and $K$ is the extrinsic curvature. After moving to imaginary time, $t = i\tau$, we can write the partition function as $Z = \exp(-I_\text{tot})$ and we can extract the entropy via
\begin{align}
S = \beta^2 \frac{\partial}{\partial \beta} \left( -\frac{1}{\beta} \ln Z \right) = \beta^2 \frac{\partial}{\partial \beta} \left( \frac{1}{\beta} I_\text{tot} \right) \, .
\end{align}
Here, $\beta$ denotes the periodicity of imaginary time that is required to avoid a conical angle deficit at the location of the black hole horizon.

\subsection{Results in General Relativity}
For a Euclideanized Schwarzschild black hole of mass $M$ one finds
\begin{align}
\beta = 8\pi G M \equiv \frac{1}{T} \, ,
\end{align}
where $T$ is the temperature associated with this system. Moreover, because the Euclideanized Schwarzschild metric is a vacuum solution, one has $I_\text{EH} \equiv 0$, and hence the black hole entropy is generated exclusively by $I_\text{GHY}$. After subtracting a reference flat space background (which can be performed at arbitrarily large distance $r \gg 2GM$ away from the black hole)  one finally arrives at
\begin{align}
S = \frac{4\pi(2GM)^2}{4\ell_\text{Pl}^2} = 4 \pi \ell_\text{Pl}^2 M^2 \, .
\end{align}
Note that the entropy $S$, the temperature $T$, and the black hole mass $M$ are not independent quantities: they are linked by the first law of black hole thermodynamics, $\dd M = T \dd S$, which is required for a consistent thermodynamic interpretation.

Keeping these considerations in mind, let us now estimate what happens if we deviate from the Euclideanized Schwarzschild geometry deep inside the black hole.

\subsection{Estimates beyond General Relativity}

The presence of a Euclidean core, as per assumption B, does not influence the location of the black hole horizon, nor its Hawking temperature; therefore, the entropy contribution of the Gibbons--Hawking--York boundary term coincides with that of General Relativity precisely. Hence, all deviations from General Relativity have to be captured by the on-shell value of $I_\text{EH}$.

To keep our considerations as agnostic as possible, we need to form a regulator distance scale $\ell > 0$ that takes into account both the Planck scale $\ell_\text{Pl}$ as well as the black hole Schwarzschild radius, $2GM$. Let us hence define, in accordance with assumption C,
\begin{align}
\ell \equiv \ell_\text{Pl}^\alpha \, (2GM)^{1-\alpha} \, ,
\end{align}
for $\alpha$ a real, continuous parameter. The appearance of such a regulator may strike the reader as odd, due to the strongly non-linear appearance of $M$, but in the literature on regular black holes such length scales are fairly well established \cite{Frolov:2016pav}. In particular, one often finds the choice $\alpha=1$ since it leads to uniformly bound curvature (in what is called the limiting curvature conjecture), but we will see that while this choice is sufficient to guarantee a consistent first law of thermodynamics (subject to assumption B), at leading order it is not necessary.

Turning to the on-shell evaluation of the Euclideanized Einstein--Hilbert action, we now estimate that the curvature is spatially confined to a Euclidean ball of the size of the regulator scale $\ell$, and that it takes on an average value of 
\begin{align}
R \sim \frac{1}{\ell^2} \, .
\end{align}
Then, for the full on-shell action---aided by assumption A---one obtains
\begin{align}
I_\text{EH} = \frac{1}{16\pi G} \int\limits_\mathcal{M} \dd^4 X_E \sqrt{g_E} R ~\sim~ \frac{1}{16\pi\ell_\text{Pl}^2} \times \beta \times \ell^3 \times \frac{1}{\ell^2} ~=~ \frac{M \ell}{2} \, .
\end{align}
The entropy contribution $\delta S$ is then
\begin{align}
\delta S = \beta^2 \frac{\partial}{\partial \beta} \left( \frac{1}{\beta} I_\text{tot} \right)
\sim \frac{M^2}{2} \frac{\partial \ell}{\partial M} = \frac{1-\alpha}{4} \left( \frac{\ell_\text{Pl}}{2GM}\right)^{\alpha-2} \, .
\end{align}
Clearly, for $\alpha > 2$, the entropy correction is suppressed, $\delta S \ll 1$. Let us now check the consistency requirement that the first law of thermodynamics is still satisfied. In order to have $\dd M = T \dd (S + \delta S)$ (since both $T$ and $M$ are invariant, as per assumption B), one needs requires $\dd \delta S = 0$. This holds if $\alpha = 1$ or $\alpha = 2$. (We could also give up assumption B, and compute the necessary $\delta T$.)

\subsection{What about higher-order operators?}
Let us now briefly consider next-to-leading order corrections to the black hole entropy stemming from higher-order actions of the form
\begin{align}
I_\text{NLO} = -\frac{1}{16\pi \, \ell_\text{Pl}^{2(2-d-r)}} \, \sum\limits_p \int\limits_{\mathcal{M}} \dd^4 X \sqrt{-g} \, R^{r-p} \, \Box^d \, R^p \, ,
\end{align}
where $2d$ counts the number of derivatives, $r$ the powers of curvature tensors, and $\sum_p$ sums all possible terns at that order. Following similar arguments above, we estimate for the on-shell value
\begin{align}
I_\text{NLO} \sim \frac{1}{16\pi \, \ell_\text{Pl}^{2(2-d-r)}} \times \beta \times \ell^3 \times \ell^{-2d} \times \ell^{-2r} = \frac{M \, \ell^{3-2(d+r)}}{2 \, \ell_\text{Pl}^{2(1-d-r)}} \, .
\end{align}
In general, however, this results in the entropy corrections
\begin{align}
\delta S = \frac{(1-\alpha)[3-2(d+r)]}{4} \left( \frac{\ell_\text{Pl}}{2GM} \right)^{2(d+r-2)(1-\alpha) - \alpha} \, .
\end{align}
As a quick cross-check, for $r=1$ and $d=0$ we recover the previous results. Compatibility with the first law is only achieved if $\alpha=1$ (because then the above term vanishes identically), or, if $2(d+r-2)(1-\alpha) - \alpha = 0$. Inserting $\alpha=2$ into that relation one finds the constraint $d+r=1$, that is, only curvature terms of the following form are allowed:
\begin{align}
I^\text{allowed}_\text{NLO} = \frac{1}{16\pi \, \ell_\text{Pl}^2} \, \int\limits_{\mathcal{M}} \dd^4 X \sqrt{-g} \, R^{r-p} \, \Box^{1-r} \, R^p \, .
\end{align}
Observe the appearance of negative powers of the d'Alembert operator for $r > 1$, similar to the nonlocal model considered by Deser and Woodard \cite{Deser:2007jk} in the context of cosmology. If one wants to allow all possible curvature-derivative couplings, however, only $\alpha=1$ remains permissible.

Alternatively, one could again modify assumption B and stipulate that the Hawking temperature changes by a contribution
\begin{align}
\frac{\delta T}{T} = - \frac{\delta \dd S}{\dd S} \, .
\end{align}
This would, however, shift the imaginary time integration domain from $\beta$ to $\beta - \delta T/T^2$, and then again affect the on-shell value of the action, leading to an iterative procedure that could be used to determine these effects more accurately (that is, if this iteration terminates reliably).

\section{Conclusions}

Replacing the singularity of a Euclidean black hole by a regular core of radius $~\ell_\text{Pl}^\alpha\,(2GM)^{1-\alpha}$ leads to corrections of the black hole entropy. If the area law and the first law of thermodynamics are to be left unchanged, at leading order this demands $\alpha=1$, or, remarkably, $\alpha=2$. In the latter case, however, higher-order corrections will again spoil the first law, which leaves $\alpha=1$ as the only option, unless one is either prepared to consider a restricted higher-order sector, similar to a nonlocal Deser--Woodard model, or one is prepared to accept a small shift in the Hawking temperature (more on that below).

The choice $\alpha=1$ sets all entropy corrections to zero, and it is related to a mean curvature of
\begin{align}
R \sim \frac{1}{\ell_\text{Pl}^2}
\end{align}
at the center of the core. Interestingly, this closely resembles the limiting curvature conjecture proposed independently by Markov and Polchinski \cite{Markov:1982,Markov:1984,Polchinski:1989}, which states that the maximum value of curvature should be uniformly bounded by the constants of nature, and be independent of the black hole mass. The choice $\alpha=2$ is compatible with the limiting curvature principle, too, if the smallest possible black hole mass is given by $M_\text{Pl}$.

But perhaps one should not disregard $\alpha=2$ too prematurely: classicalization, for example, is a mechanism that renders gravity classical at smaller scales \cite{Dvali:2010jz,Dvali:2016ovn}, and could hence result in a bound state of soft gravitons at the black hole center. Viewed from that perspective, quadratic curvature terms may not be necessary to understand gravity in these regimes. Moreover, the first law of black hole thermodynamics can be recovered if the Hawking temperature itself receives a small correction $\delta T = -T \delta \dd S/\dd S$, which is proportional to the change in entropy.

At any rate, more accurate calculations are needed to compute the entropy corrections from first principles, but we hope that this Essay can serve as a motivation to think more about the role of the black hole center and its possible contributions to the Euclidean gravitational path integral.

\section*{Acknowledgements}
I thank the NSF for support under Grants PHY-1819575 and PHY-2112460.

\begin{singlespace}

\end{singlespace}

\end{document}